\documentclass[12pt]{article}

\usepackage{epsfig,graphicx}
\usepackage{amssymb,amsmath,amsthm}

\title{ Low-energy effects in brane worlds: Liennard-Wiechert potentials and Hydrogen Lamb shift\footnote{We are pleased
to dedicate this work to Professor Octavio Obreg\'on on occasion of his 60th birthday. }}
\author{ Hugo A. Morales-T\'ecotl\thanks{Associate member of the Abdus Salam ICTP, Trieste Italy.}, Omar
Pedraza and Luis O. Pimentel\\
Departamento de F\'{\i}sica \\Universidad Aut\'onoma Metropolitana Iztapalapa \\
San Rafael Atlixco 186, CP 09340, M\'exico D.F., M\'exico. \\
E-mail: {\tt hugo,omp,lopr@xanum.uam.mx}}

\begin{document}

\maketitle

\begin{abstract}
\small Testing extra dimensions at low-energies may lead to interesting effects. In this work a test point
charge is taken to move uniformly in the 3-dimensional subspace of a (3+$n$)-brane embedded in a (3+$n$+1)-space
with $n$ compact and one warped infinite spatial extra dimensions. We found that the electromagnetic potentials
of the point charge match standard Liennard-Wiechert's at large distances but differ from them close to it.
These are finite at the position of the charge and produce finite self-energies. We also studied a localized
Hydrogen atom and take the deviation from the standard Coulomb potential as a perturbation. This produces a Lamb
shift that is compared with known experimental data to set bounds for the parameter of the model. This work
provides details and
extends results reported in a previous Letter.
\end{abstract}

\baselineskip=20pt

\section{Introduction}

Spacetimes with more than three spatial dimensions have been attracting interest for long time. First proposals
go back to  Kaluza and Klein \cite{kk}, in their attempt to unify electromagnetism with Einstein gravity by
proposing a theory with a compact fifth dimension of a size of the order of the Planck length. Later on the
emphasis shifted to a ``brane world" picture \cite{Akama, Rubakov, Visser} with our world being confined to a
subspace of higher-dimensional spacetime. More recently it was put forward, inspired also in \cite{HEI,HEII},
that at energy scales of the Standard Model matter cannot propagate into extra dimensions whilst gravity on the
other hand can permeate all over. Within brane worlds  extra dimensions may be compact but large
\cite{Arkani-Hamed:1998rs, AADD, ADD}, or infinite and warped \cite{Randall:1999ee,Randall:1999vf}. They have
opened a rich and interesting phenomenological model-building aiming at solving long standing problems like the
hierarchy and cosmological constant problem  within high-energy physics \cite{Rubakov:2001kp, feruglio,
Perez-Lorenzana:2005iv}, as well as testing models with observational data at the cosmological level \cite{roy}.

We should stress at this point that to implement the brane scenario  one looks for  physical mechanisms which
localize higher dimensional fields to a lower dimensional brane. Localization of gauge fields has turned out to
be more difficult than that of scalar and spinors, however, some mechanisms have been developed in
\cite{Dubovsky:2000av, Neronov:2001br, giovaninni, RD-Sh}. The one in \cite{Dubovsky:2000av} is a particularly
simple scenario  that was probed to localize electromagnetic fields. It consists of a single (3+$n$)-brane
embedded in a (3+$n$+1)-dimensional space containing $n$ compact extra dimensions and one infinite and warped.
This is a (5+$n$)-dimensional brane world that we refer to as a Randall-Sundrum II model modified by $n$ extra
compact dimensions (RSIIn). As for fermions some mechanisms have been developed in \cite{Jackiw:1975fn,
Bajc:1999mh, Dubovsky:2000am, Neronov:2001qv, spinor}. We shall combine some of these ideas for our purposes
below.

It is generally assumed within brane models that relevant effects appear at high energies \cite{Rubakov:2001kp,
feruglio, Perez-Lorenzana:2005iv} whereas low-energy effects have no chance to become relevant. This is not
necessarily the case since some high-precision tests may be sensible to such tiny effects or else, at worst, it
may be possible to set bounds for the different parameters of the models \cite{Bluhm1, Bluhm2, LMP-Casimir}. In
this paper we investigate two low-energy effects for RSIIn. The first one involves determining the
electromagnetic potentials of point charge in uniform motion in a 3-dimensional subspace of the single
(3+$n$)-brane. Considering that experiments in atomic physics are sensitive to small frecuency shifts down to 1
mHz \cite{Bluhm1,Bluhm2}, one can in principle use the atom as an instrument to measure the anti-de Sitter
radius $\epsilon$. The second problem we investigate is a Hydrogen atom, localized on the brane, where the
binding electromagnetic field is just the one obtained in the first part. We have calculated the
$2S_{1/2}-2P_{1/2}$ splitting coming from the presence of the extra dimensions that corresponds to the Lamb
shift associated to QED. This has allowed us to set bounds for the parameter of the scenario, namely the anti-de
Sitter radius or, equivalently, the tension of the brane. In the present work we provide details and extend some
of the results of a previous Letter \cite{Letter}.

The paper is organized as follows. In section II we describe the RSIIn brane world and describe the localization
mechanism for the electromagnetic and spinor fields. Section III is devoted to calculate the effective
electromagnetic potentials for a test point charge in uniform motion thus providing details and generalizing the
static results of \cite{Letter}. In section IV the $2S_{1/2}-2P_{1/2}$  shift of a localized Hydrogen atom is
calculated completely. Finally in section V we focus on the discussion of our results as well as some
perspectives.
\section{The scenario RSIIn}
Now we describe the brane world we use in the present work. Looking for a localization mechanism for gauge
fields as due to the gravity associated with the brane it was proved in \cite{Dubovsky:2000av} that the
following setting would accomplish such task: A single (3+$n$)-brane embedded in a (3+$n$+1)-dimensional space
so that space-time is ($5+n$)dimensional with $n$ compact dimensions and one infinite and warped. We call such a
scenario RSIIn. Much as in the case of the Randall-Sundrum models to solve Einstein equations the brane tension
and the negative bulk cosmological constant must be fine-tuned. The resulting metric is \cite{Dubovsky:2000av}
\begin{equation}\label{ecu:fon}
ds^2 = a(y)^2\left(\eta_{\mu\nu}dx^{\mu}dx^{\nu} -\sum_{i=1}^nRd\theta_i^2\right)-dy^2,\qquad
a(y)=e^{-\frac{|y|}{\epsilon}}\,,
\end{equation}
where $\eta_{\mu\nu}$ is the four-dimensional Minkowski metric, $\theta_i$ $\in$ $[0,2\pi]$ are compact
coordinates and $R_i$ are the sizes of the compact dimensions. $\epsilon$ is the anti-de Sitter radius.

Since our purpose is to study low energy-physics effects and hence the situation where electromagnetic fields
and spinors are localized to a 3-dimensional subspace of the (3+$n$)-brane we next describe how this works in
RSIIn.

\subsection{Localization of the electromagnetic field}
There has been extensive discussion about the localization of gauge fields in brane-worlds
\cite{Dubovsky:2000av, Neronov:2001br, RD-Sh}. Here we just reproduce the proposal in \cite{Dubovsky:2000av} for
the  RSIIn scenario. Consider a gauge field $A_M$, with action
\begin{equation}
S_g=-\frac{1}{4}\int dx^4 dy\prod \frac{d\theta_i}{2\pi R_i} \sqrt{g}g^{MP}g^{NQ}F_{MN}F_{PQ} \,.
\end{equation}
For low-energies, hence  $R_i$ small, and in the background (\ref{ecu:fon}), we truncate this action to the zero
Kaluza-Klein modes of the compact $n$ dimensions. The effective 4-dimensional electromagnetic action turns out
to be
\begin{equation}
S=-\frac{1}{4}\int_{-\infty}^{\infty} dy\,e^{-n\frac{|y|}{\epsilon}}\int
dx^4\eta^{\mu\nu}\eta^{\lambda\rho}F_{\mu\lambda}F_{\nu\rho} \,.
\end{equation}
Hence $A_{\mu}$ is normalizable, up to a normalization factor proportional to $\sqrt{\frac{n}{\epsilon}}$. If
$n=0$ there is no localized gauge field, but for $n\geq 1$ the gauge field localizes on the brane and
electrodynamics on it becomes 4-dimensional at large distances $l>\!\!> R_i$. We proceed next to discuss the
spinor localization in RSIIn.

\subsection{Localization of spinors}
As in the case of the electromagnetic field in RSIIn, the observed 4-dimensional fields are the zero modes of
higher dimensional ones. For explicit calculations we shall be considering $n$=1,2, namely 6- and 7-dimensional
spacetime unless we specify otherwise, and hence 8-component spinors. The analysis should hold with the adequate
generalizations for arbitrary $n$. We combine the localization mechanism of \cite{Dubovsky:2000am} designed to
work in 5-dimensional spacetime for massive fermions with the projection analysis of
\cite{Neronov:2001qv,Randjbar-Daemi:1982hi} that allows one to identify 4-component spinors out of the higher
dimensional ones.

To begin with a domain wall is considered to be produced by some scalar field $\chi$. The domain wall separates
two regions  in the non-compact extra dimension $y$. Explicitly the region $\chi=-v$, $y<0$ separates from
$\chi=v$, $y>0$. One introduces a Yukawa coupling between the spinor and the scalar $g\chi \bar\psi\psi$. For a
given sign of the Yukawa coupling $g$ the zero modes have chirality defined. In general there is no dependence
on the details of the profile of the scalar field across the wall. It turns out to be convenient to organize the
chiral spinors into one field $\Psi$. The mass term is given by $\mu\bar\Psi\tau_1\Psi$. At low-energies, and
hence small $R_i$, we truncate this action to the zero Kaluza-Klein modes of the compact $n$ dimensions. The
resulting fermion action reads, after an adequate internal rotation \cite{Dubovsky:2000av},
\begin{eqnarray}
S&=&\int dxdy\sqrt{g}\bar\Psi\left(\Gamma^AE_A^M\left[\partial_M+\Omega_M+ie^{(n)}A_M\right]+g\chi
\tau_1+\mu\tau_2\right)\Psi, \\
\Psi &=& \left(\begin{array}{c} {\psi_1} \\ {\psi_2} \end{array}\right) \,.
\end{eqnarray}
Here $\psi_1$ and  $\psi_2$ are spinors in ($5+n$)-dimensions and
$\Omega_M=\frac{1}{2}\Omega_{M[AB]}\Sigma^{AB}$ is the spin conection, with generators
$\Sigma_{AB}=\frac{1}{4}[\Gamma_A,\Gamma_B]$. In what follows we restrict to the cases $n=1,2$ . In this way
$\Gamma^A$ are six $8\times 8$ Dirac matrices. The Dirac equation which follows from this action in the
background (\ref{ecu:fon}) has the form
\begin{equation}\label{ecu:dir1}
\left[\Gamma^{A}E_A^M\left\{\partial_{M}+\Omega_M+ie^{(n)}A_M\right\}-g\chi(y)\tau_1-\mu\tau_2\right]\Psi=0 \,.
\end{equation}
Since the details of the profile of the scalar filed across the wall play no role for the localization of
fields, we use an infinitely thin wall approximation. In this limit one has $g\chi(y)=gv$sign$(y)$. The
components $\Omega_A$ are given in terms of the frames components through $\Omega_A=E_A^M\Omega_M$. Where the
only non zero frames components are
\begin{equation}
E_a^{\mu}=\frac{1}{a(y)}\delta^{\mu}_a,\qquad E^y_{\underline{y}}=1,\quad
E^{\theta_1}_{\underline{\theta_1}}=\frac{1}{a(y)},\qquad E^{\theta_2}_{\underline{\theta_2}}=\frac{1}{a(y)} \,.
\end{equation}
With  $a=0,1,2,3$, the indices in the coordinate $\theta$  refer to $n=1,2$ and underlined indices refer to the
orthonormal frame. The non-vanishing ones are
\begin{equation}
\Omega_{\mu}=\frac{1}{2}a(y)'\Gamma_{y}\Gamma_{\mu},\quad\Omega_{\theta_1}=\frac{1}{2}a(y)'
\Gamma_y\Gamma_{\theta},\quad\Omega_{\theta_2}=\frac{1}{2}a(y)'\Gamma_y\Gamma_{\theta}\,.
\end{equation}
When  $\mu\neq 0$ and $gv> 1/2\epsilon$ there are no true localized modes but a metastable massive state
\cite{Bajc:1999mh,Dubovsky:2000am}. We assume such conditions hold for our $\chi$ and (\ref{ecu:dir1}) becomes
\begin{equation}\label{ecu:dir2} \left[\frac{\Gamma^{\mu}\{\partial_{\mu}+e^{(n)}A_{\mu}\}}{a(y)}+ \Gamma^y\partial_{y}
-g\chi(y)\tau_1-\mu\tau_2\right]\Psi=0 \,. \end{equation} We proceed now by choosing Dirac matrices in the
following form: $\Gamma_{\mu}=\gamma_{\mu}\otimes\sigma_3$, $\Gamma_{y}=\gamma_{5}\otimes\sigma_3$ and
$\Gamma_{\theta}=1\otimes\sigma_2$. A convenient rearrangement for $\Psi$ is \cite{Neronov:2001qv}:
\begin{equation}
\Psi=\phi(x)\otimes{\cal U}\otimes{\cal V}(y),\quad\Rightarrow \Gamma^{y}\otimes I_{2\times 2}\Psi_{R,L}=\pm
\Psi_{R,L},\qquad {\cal U}=\binom{1}{0} \,.
\end{equation}
Because of the presence of $\phi(x)$, from the point of view of 3-dimensional space, the spinor represents a
quantum point-like particle. Now, the fundamental representation for $\Psi$ corresponds to the eigenvalues $+1$
and $-1$ for $\tau_3$, thus we use $\tau_3\Psi=\Psi$ as in \cite{Randjbar-Daemi:1982hi}. In terms of
$\Psi_{L,R}$ the equation(\ref{ecu:dir2}) translates into a set of coupled equations,
\begin{eqnarray}
\frac{\Gamma^{\mu}\{\partial_{\mu}-e^{(n)}A_{\mu}\}}{a(y)}\Psi_L+
\partial_{y}\Psi_R
-(g\chi(y)\tau_1+\mu\tau_2)\Psi_R=0\label{ecu:1}\,,\\
\frac{\Gamma^{\mu}\{\partial_{\mu}-e^{(n)}A_{\mu}\}}{a(y)}\Psi_R-\partial_{y}\Psi_L
-(g\chi(y)\tau_1+\mu\tau_2)\Psi_L=0\label{ecu:2} \,.
\end{eqnarray}
By taking
\begin{equation}
\gamma^{\mu}\{\partial_{\mu}-e^{(n)}A_{\mu}\}\phi=m\phi
\end{equation}
and after eliminating $\Psi_L$ in the system (\ref{ecu:1})-(\ref{ecu:2}) one obtains a second order equation for
${\cal V}(y)$ that is given by
\begin{equation}
\left\{\frac{m^2}{a^2}+\partial_y^2-g\chi'\tau_1-\kappa\,\mathrm{sign}(y)\partial_y+\kappa\,\mathrm{sign}(y)(g\chi
\tau_1+\mu\tau_2)+(g\chi \tau_1+\mu\tau_2)^2\right\}{\cal V}(y)=0
\end{equation}

>From here on the same argument of \cite{Dubovsky:2000am} holds and there is a metastable state. Namely, for
$\mu<<1/\epsilon$ one has
\begin{equation} m=m_0-i\Delta
\end{equation}
with
\begin{equation}
m_0=\left(1-\frac{1}{gv\epsilon}\right)\mu,\quad\Delta=m_0\left(\frac{m_0\epsilon}{2}\right)^{2gv\epsilon-1}\,
\frac{\pi}{\left[\Gamma(gv\epsilon+1/2)\right]^2}
\end{equation}
with $\Gamma(\dots)$ being the Gamma function. In the opposite regime, $\mu>>1/\epsilon$, one has
\begin{equation}
m_0=\mu,\quad\Delta=\frac{m_0}{2}\left(\frac{m_0}{2M}\right)^{2M\epsilon-1}\,e^{2M\epsilon},\quad
M=\sqrt{(gv)^2+\mu^2}
\end{equation}
Overall, for a given scalar field fulfilling the conditions above, that is to say $gv>1/2\epsilon$, it is
possible to envisage a long life-time for the spinor to be localized on the brane. We assume this is the case
and hence the role of the extra dimensions in RSIIn at low-energies for a localized Hydrogen atom we are
interested in is reduced to an effective electromagnetic potential for a point-like  proton which we describe in
the following section.

\section{Liennard-Wiechert potentials for a point particle}
As a first step in the calculation of the energy levels corrections for Hydrogen atoms one has to find the
effective electrostatic potential that replaces  the Coulomb potential. In \cite{Letter} we have presented the
electrostatic potential produced by a point charge localized in the single brane. A uniformly moving 4D point
charge source will be considered as follows
\begin{eqnarray}
\sqrt{g}j^{\mu}&=&\sigma^{(5+n)}\ v^{\mu}\delta^{3}(\vec{ x}-\vec{\mathrm{ x}}(t))\delta(y-\mathrm{y}(t))\label{eq:jmu}\,,\\
\vec{\mathrm{x}}(t) &=& \vec{v} t \,,\\
\mathrm{y}(t) &=& \zeta \,.
\end{eqnarray}
Here $\sigma^{(5+n)}$ is the charge density and $v^{\mu}$ is the velocity of the source and $\zeta$ is a
parameter that turns out to be useful in regularizing a delta product later on so we are interested at the end
in the limit $\zeta\rightarrow 0$. For completeness we briefly recall the calculation leading to the photon
Green's function as given in the reference \cite{Dubovsky:2000av}. For the background (\ref{ecu:fon}) the
Maxwell equations on the right to the brane yield
\begin{eqnarray}
&&{\cal O}\hat A^{\sigma} -e^{-n\frac{|y|}{\epsilon}}\hat\partial^{\sigma}\partial_{\mu}\hat A^{\mu}
= -R^{-n}\sqrt{g}\,\,j^{\sigma},\label{eq:MaxRSn}\\
&&{\cal O}:=-e^{-(n+2)\frac{|y|}{\epsilon}}\partial^2_{y}+
\frac{(n+2)}{\epsilon}e^{-(n+2)\frac{|y|}{\epsilon}}\, \mathrm{sign}(y)
\partial_{y}+e^{-n\frac{|y|}{\epsilon}}\Box ,
\label{eq:OMax}
\end{eqnarray}
here we use the gauge $A^{y}=0$, $A^{\theta_i}=0$. $\hat A^{\nu}=\eta^{\nu\mu}A_{\mu}$ and $\Box$ is the 4D
dalambertian. The term containing $\partial_{\mu}\hat A^{\mu}$ is pure gauge on the brane so we just drop it
from now on. One now proceeds to solve (\ref{eq:MaxRSn}) using the Green's function approach, which satisfies
\begin{equation}
 {\cal O}G(y,y')=\delta(x-x')\delta(\theta-\theta')\delta(y-y')\,.
 \end{equation}
We solve this equation using the eigenfunction expansion for the Green's function
\begin{equation}\label{ec:green}
G(x, x',y,y') =\int \frac{d^4k}{(2\pi)^4} e^{ik(x-x')}\left[\frac{\phi_0(y)\phi_0(y')}{k^2}+\int
dm\frac{\phi_m(y)\phi_m(y')}{k^2-m^2}\right].
\end{equation}
Thus the corresponding eigenvalue equation   for $\phi$ can be rewritten as
\begin{equation}
\left[-e^{-(n+2)\frac{|y|}{\epsilon}}\partial^2_{y}+ \frac{(n+2)}{\epsilon}e^{-(n+2)\frac{|y|}{\epsilon}}\,
\mathrm{sign}(y)
\partial_{y}+e^{-n\frac{|y|}{\epsilon}}m^2\right]\phi(y)=0.\label{ecSL}
\end{equation}
Notice that this equation is invariant under a change of sign in the coordinate $y$ and therefore it is enough
to solve the equation in the region $y > 0$ (on the right to the brane). Performing the change of variable
$\xi=\epsilon e^{\frac{y}{\epsilon}}$ and redefining the function as $\phi=\xi^{\nu}\hat \phi,\quad\nu=1+n/2$,
one gets the Bessel equation
\begin{eqnarray}\label{ecu:5}
\partial_{\xi}^{2}\hat \phi_{m}
+\frac{1}{\xi}
\partial_{\xi}\hat \phi_{m}+\bigg(1-\frac{\nu^2}{\xi^2}\bigg)\hat \phi_{m}&=&0\,.
\end{eqnarray}
When $m = 0$ the solution has the form
\begin{equation}
\phi_0=\sqrt{\frac{n}{2\epsilon}}\,,\,\, \int_{-\infty}^{\infty}dy\,e^{-n\frac{|y|}{\epsilon}}\phi_0^2 =1.
\end{equation}
The weight factor $e^{-n\frac{|y|}{\epsilon}}$ comes from the Sturm-Liuville form of Eq. (\ref{ecSL}). In the
case $m > 0$ the normalized modes become
\begin{equation}
\phi_m =\sqrt{\frac{m\epsilon}{2}}\,\left(\frac{\xi}{\epsilon}\right)^{\nu}\left( \frac{
N_{\nu-1}\left(m\epsilon\right)J_{\nu}\left(m\xi\right) -J_{\nu-1}\left(m\epsilon\right)N_{\nu}\left(m\xi\right)
}{\sqrt{N^2_{\nu-1}\left(m\epsilon\right)+J^2_{\nu-1}\left(m\epsilon\right)}} \right),
\end{equation}
fulfilling the boundary condition $\partial_y\phi_m(y)\big|_{y=0}=0$ as well as the normalization
\begin{equation}\label{ecu:7}
\int_{-\infty}^{\infty}dy e^{-n\frac{|y|}{\epsilon}}\phi_m(m\epsilon)\phi_{m'}(m'\epsilon)=\delta(m-m')\,.
\end{equation}
Performing the integration over $k$ in Eq.(\ref{ec:green}) yields the Green's function on the brane
\begin{eqnarray}\label{ec:green1}
G_R(x-x',y=0,y') &=& \frac{n}{2\epsilon} D_0(x-x') +
\int dm \phi_m(0) \phi_m(y') D_m(x-x'), \label{eq:Green}\\
D_0(x-x') &=& \frac{\delta(\tau-r)}{4\pi r},\ \tau:= t-t',
\ r:=|\vec{x}-\vec{x}'|,\label{D0}\\
D_m(x-x') &=&D_0(x-x')+\frac{\theta(\tau-r)} {4\pi\sqrt{\tau^2-r^2}}mJ_1(m\sqrt{\tau^2-r^2}).\label{eq:Dm}
\end{eqnarray}
Here the first term is the contribution of the zero mode, $m=0$, the second term comes from the continuum of
massive modes, $m>0$. The induced electromagnetic potential is obtained as usual by integrating the Green's
function with the source which in this case becomes
\begin{equation}
A^{\mu}(x,y=0)=-\int d^4 x'dy' dm\frac{\theta[\tau-r]}{4\pi\lambda}mJ_1(m\lambda)  \phi_m (0)\phi_m
(y')R^{-n}\sqrt{g}\,\,j^{\mu}(x',y')\,. \label{eq:Ajmu}
\end{equation}
We have used  equations (\ref{eq:Green})-(\ref{eq:Dm}), $\lambda:=\sqrt{\tau^2-r^2}$  and the completeness
relation
\begin{equation}
\phi_0(y)\phi_0(y') + \int dm \phi_m(y) \phi_m(y') = \delta(y-y')\,.
\end{equation}
Notice that the eq. (\ref{eq:Ajmu}) is obtained by regularization of the product of $\delta$'s as follows:
assume $\mathrm{y}(t)=\zeta$, with $\zeta$ a regulating parameter going to zero in the limit. This yields zero
because of the delta product.

Now we focus on the evaluation (\ref{eq:Ajmu}). For low energies only small masses $m$ ($m<\!\!<1/\epsilon$) are
relevant, and we obtain for the electromagnetic potential
\begin{equation}\label{eq:Amur}
A^{\mu}=-\frac{1}{R^n}\int dt' dm\frac{v^{\mu}\theta[(t-t')-|\vec{x}-\vec{\mathrm{ x}}(t')|]}{ 4\pi
\Gamma(\nu-1) }  \left(\frac{\epsilon}{2}\right)^{\nu-1} \sigma^{(5+n)}m^{\nu} J_{\nu}\left(\epsilon
m\right)\frac{J_1\left(m \lambda\right)}{\lambda}\,.
\end{equation}
For the evaluation of  equation (\ref{eq:Amur}), there are two cases, whether ${n}$ is even or odd. We evaluate
first the equation (\ref{eq:Amur}) for $n$ odd ($\nu$ half-integer). We  use the relation
\begin{equation}\label{ecu:re2}
m^{l+1/2}J_{l+1/2}\left(m\epsilon\right)=\left(-1\right)^{l}\sqrt{\frac{2}{\pi}}\epsilon^{l+1/2}\left(\frac{d}{\epsilon
d\epsilon}\right)^{l}\frac{\sin(m\epsilon)}{\epsilon},
\end{equation}
and introduce (\ref{ecu:re2}) in (\ref{eq:Amur}) to integrate over $m$. So we easily find the result
\begin{equation}\label{eq:poestn}
A^{\mu}=\sqrt{\frac{2}{\pi}}\int_{-\infty}^{t_*} dt' \frac{(-1)^{\nu-1/2}\theta[(t-t')-|\vec{x}-\vec{\mathrm{
x}}(t')|]}{4\pi R^n \Gamma(\nu-1)\lambda}
\sigma^{(5+n)}v^{\mu}\frac{\epsilon^{2\nu-1}}{2^{\nu-1}}\left(\frac{d}{\epsilon d\epsilon}\right)^{\nu-1/2}
\frac{\theta[\lambda-\epsilon]}{\lambda\sqrt{\lambda^2-\epsilon^2}},
\end{equation}
where $t_*$ satisfies $(t-t_*)^2=\epsilon^2+|\vec x-\vec{\mathrm{ x}}(t_*)|^2$. It is possible to calculate the
potential when the particle moves with uniform speed. We consider, for simplicity, the case when the motion
takes place along the $x$ direction in the $xy$ plane, the solution is

\begin{equation}
A^0=\sqrt{\frac{2}{\pi}}\frac{(-1)^{\nu-1/2}\sigma^{(5+n)}\epsilon^{\nu}}{4\pi R^n\Gamma(\nu-1)}
\frac{\epsilon^{2\nu-1}}{2^{\nu-1}} \left(\frac{d}{\epsilon d\epsilon}\right)^{\nu-1/2}
\frac{\arctan{\left(\frac{\sqrt{y^2(1-v^2)+(x-vt)^2}}{\epsilon\sqrt{1-v^2}}\right)}}{\epsilon\sqrt{y^2(1-v^2)+(x-vt)^2}}.
\end{equation}

For a static particle we have
\begin{equation}\label{ecu:pot1}
A^0=\sqrt{\frac{2}{\pi}}\frac{(-1)^{\nu-1/2}}{4\pi
R^n\Gamma(\nu-1)}\sigma^{(5+n)}\frac{\epsilon^{2\nu-1}}{2^{\nu-1}}\left(\frac{d}{\epsilon
d\epsilon}\right)^{\nu-1/2}\frac{1}{2r\epsilon}\left[\arctan{\left(\frac{r}{\epsilon}\right)}\right],\quad
r=|\vec{x}|.
\end{equation}

For the case of $n$ even ($\nu$ integer), we can use the relation
\begin{equation}\label{eq:npar}
m^{\nu}J_{\nu}\left(m\epsilon\right)=(-1)^{\nu}\epsilon^{\nu}\left(\frac{d}{\epsilon d\epsilon}\right)^{\nu}J_0
\left(m\epsilon\right)
\end{equation}
and integrating over $m$ the electromagnetic potential can be rewritten as
\begin{equation}\label{eq:mures2}
A^{\mu}=-\int_{-\infty}^{t_*} dt'\frac{\theta[(t-t')-|\vec{x}-\vec{\mathrm{ x}}(t')|]}{4\pi R^n\Gamma(\nu-1)}
\frac{\sigma^{(5+n)}v^{\mu}(-1)^{\nu}}{\lambda}\frac{\epsilon^{2\nu-1}}{2^{\nu-1}} \left(\frac{d}{\epsilon
d\epsilon}\right)^{\nu-1}\frac{\delta[\lambda-\epsilon]} {\epsilon\lambda}.
\end{equation}

The electromagnetic potential for a point particle that moves with uniform speed  is given by
\begin{equation}
A^{\mu}=-\frac{(-1)^{\nu+1}\epsilon^{2\nu}}{4\pi R^n \Gamma(\nu-1)2^{\nu-1}}{\left(\frac{d}{\epsilon
d\epsilon}\right)}^{\nu-1} \frac{\sigma^{(5+n)}v^{\mu}}{\epsilon^2(\sqrt{\epsilon^2+|\vec{x}-\vec{\mathrm{
x}}(t_*)|^2}-\hat n\cdot \vec{v}|\vec{x}-\vec{\mathrm{ x}}(t_*)|)}\,,
\end{equation}
where $\hat n=[\vec{x}-\vec{\mathrm{ x}}(t_*)]/[|\vec{x}-\vec{\mathrm{ x}}(t_*)|]$.

For the static particle we have the following potential
\begin{equation}\label{ecu:pot2}
A^{0}=\frac{ (-1)^{\nu+1}  }{4\pi R^n\Gamma(\nu-1)} \frac{\epsilon^{2\nu-1}}{2^{\nu-1}}\left(\frac{d}{\epsilon
d\epsilon}\right)^{\nu-1}  \frac{\sigma^{(5+n)}\theta
(\sqrt{\epsilon^2+r^2}-r^2)}{\epsilon^2\sqrt{\epsilon^2+r^2}},\quad r=|\vec{x}|.
\end{equation}

Let us present the above results for $A^0(x, t)=A^0(r, t)$ for $n = 1$ and $n = 2$.

$n=1$:\\ Liennard-Wiechert\\
\begin{equation}
A^0=\frac{e^{(n=1)}_{\mathrm{eff}}}{\epsilon}\left\{
\frac{1}{\sqrt{1-v^2}\left(1+\frac{y^2(1-v^2)+(x-vt)^2}{\epsilon^2(1-v^2)}\right)}+
\frac{\arctan{\left(\frac{\sqrt{y^2(1-v^2)+(x-vt)^2}}{\epsilon\sqrt{1-v^2}}\right)}}{\frac{\sqrt{y^2(1-v^2)+(x-vt)^2}}{\epsilon}}
\right\},\label{ecua:pim}
\end{equation}
Static case\\
\begin{equation}
A^0=\frac{e^{(n=1)}_{\mathrm{eff}}}{\epsilon}\left\{\frac{1}{1+\tilde x^2}+\cfrac{\arctan\left(\tilde
x\right)}{\tilde x}\right\},\quad \tilde x=\frac{r}{\epsilon}, \quad
e^{(n=1)}_{\mathrm{eff}}:=\frac{\sigma^{(6)}}{2R\pi \epsilon^2}.\label{ecu:po1}
\end{equation}

$n=2$:\\ Liennard-Wiechert
\begin{eqnarray} A^{0}&=&\frac{ e^{(n=2)}_{\mathrm{eff}} }{\epsilon } \left\{
\frac{1}{\left[\sqrt{1+R(t_*)^2}-\hat n\cdot \vec{v}R(t_*)\right]} + \frac{\left(1-|\vec
v|^2\right)}{2\left[\sqrt{1+R(t_*)^2}- \hat n\cdot \vec{v}R(t_*)\right]^3} \right\},\label{ecua:lw}\\
R(t_*)&=&\frac{|\vec{x}-\vec{\mathrm{ x}}(t_*)|}{\epsilon}
\end{eqnarray}
Static case \begin{equation}A^0=\frac{e^{(n=2)}_{\mathrm{eff}}}{\epsilon}\left\{\frac{1}{\sqrt{1+\tilde
x^2}}+\frac{1}{2\left(1+\tilde x^2\right)^{3/2}} \right\},\quad
e^{(n=2)}_{\mathrm{eff}}:=\frac{\sigma^{(7)}}{R^2\pi \epsilon}. \label{ecu:po2}
\end{equation}

\begin{figure}
\center{\includegraphics[height=10cm,angle=-90]{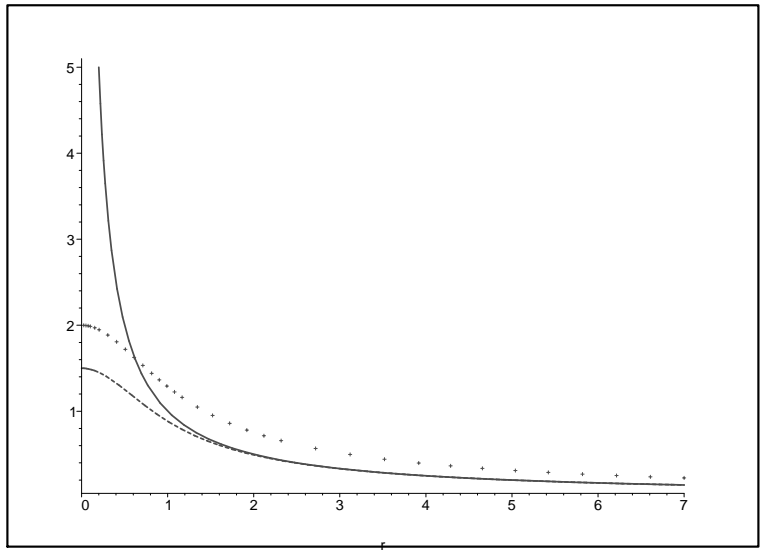}} \caption{Effective electrostatic potential for
the point particle multiplied, respectively, by the factor $4\pi\epsilon_0/e$ for the standard 4D case and
$4\pi\epsilon_0/e_{eff}$   for $n=1,2$ compact extra dimensions. $e_{eff}$ is the effective charge of the
particle in the latter case. The continuous curve represents standard 4D Coulomb potential, the crosses
correspond to $n=1$  whereas the dashes are associated with $n=2$. Units are such that $\epsilon=1$.}
\label{Coulomb}
\end{figure}

Our electrostatic potentials for $n=1,2$ are consistent with the Coulomb potential at long distances and they
preserve spherical symmetry. Remarkably these electrostatic potentials are finite at the position of the charge.
See Fig. (\ref{Coulomb}). The finite character of our effective electrostatics potentials allows to maintain the
idea of an effective 3D point particle.

\begin{figure}
\center{\includegraphics[height=10cm,angle=-90]{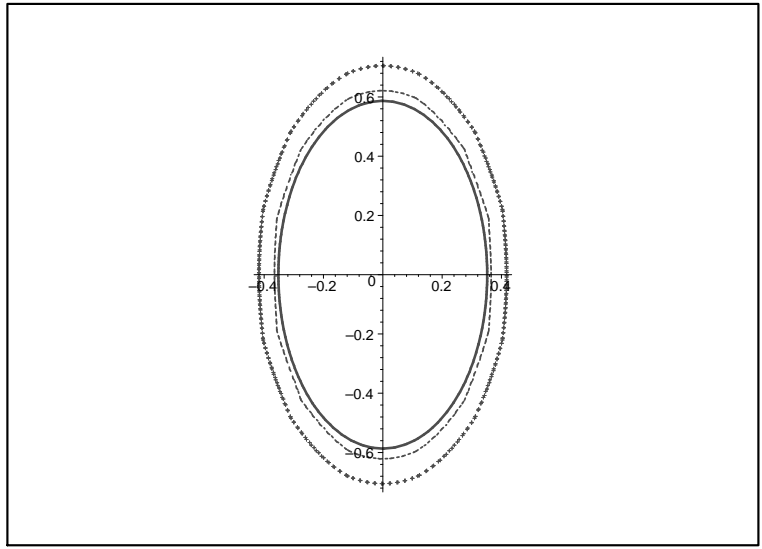}} \caption{Equipotential surfaces of the effective
Liennard-Wiechert potentials for the point particle. The continuous curve represents the standard 4D case, the
dashes $n=1$ and the crosses $n=2$. A value of the potential of 2.98 was used for the potential multiplied by
the factor $4\pi\epsilon_0/e$ for the standard 4D case and $4\pi\epsilon_0/e_{eff}$ with $e_{eff}$ for the
effective charge of the particle for $n=1,2$ compact extra dimensions. Units are such that $\epsilon=1$ and the
speed of the particle is $v=0.9$ along the horizontal axis.} \label{lwp}
\end{figure}

Fig. (\ref{lwp}) shows the  effective electromagnetic and standard 4D Liennard-Wiechert potentials. In polar
coordinates with the origin at the particle's position at the present time and $R(t_*)$ given by the solution of
\begin{equation}
R^2(1- v^2)+ 2v\rho\sqrt{\epsilon^2+R^2}\,\cos\theta- \rho^2 = 0.
\end{equation}
So $R$ can take on the values
\begin{eqnarray}
R_{1,2}&=&\frac{\sqrt{\rho^2(2v^2\cos^2\theta-v^2+1)\pm 2v\rho\Sigma \cos\theta }}{1-v^2},\\
\Sigma &:=&\sqrt{\rho^2(v^2\cos^2\theta-v^2+1)+\epsilon^2(1-v^2)^2}
\end{eqnarray}
In the electromagnetic potential Eq. (\ref{ecua:lw}) $R(t_*)= R_1$ for $\theta \in[\pi/2,3\pi/2]$ and
$R(t_*)=R_2$ for $\theta \in[-\pi/2,\pi/2]$.

\section{Hydrogen Lamb shift} In this section we calculate the possible corrections to the
energy levels for a Hydrogen atom caused by the existence of extra dimensions in RSIIn. The effective
four-dimensional   Dirac equation is given by
\begin{equation}\label{ecu:dira}
\left(i\gamma^{\mu}\partial_{\mu}-e^{(n)}\gamma^0A_0-m\right)\phi=0.
\end{equation}
Here $A^0$ is given by (\ref{ecu:po1}) in the case of  one compact extra dimension and (\ref{ecu:po2}) for two
compact extra dimensions and $e^{(n)}$ stands for the electron charge as coming from $5+n$ dimensional
spacetime. Given the complexity of the effective electrostatic potential we calculate
 the  energy shifts by means of perturbation theory to first
order.

We consider first the $n=1$ compact extra dimension case. Expanding
 the electrostatic potential (\ref{ecu:po1}) around the Coulomb potential, {\em ie} $r>\!\!>\epsilon$, gives
\begin{equation}
e^{(n=1)}A_0\approx-\alpha\left\{\frac{1}{r}-\frac{4}{3\pi} \frac{\epsilon^3}{r^4}\right\}\, ,\quad \alpha:=
e^{(n=1)}_{{\mathrm{ eff}}}e^{(n=1)}.
\end{equation}
>From here on we shall assume $\alpha=\alpha_{\mathrm{QED}}$ in order to compare with experimental data. Within
the present approximation the stationary part of Dirac equation (\ref{ecu:dira}) can be rewritten as
\begin{eqnarray}
E\phi&=&({\mathbf H}_D+{\mathbf H})\phi,\\
{\mathbf H}_D&=&{\boldsymbol\alpha}\cdot {\mathbf p}+
m{\boldsymbol\beta}-\frac{\alpha}{r},\\
{\mathbf H}^{(n=1)}&=&\frac{4\alpha}{3}\frac{\epsilon^3}{r^4},
\end{eqnarray}
where the eigenvalue  $E$ is the energy and ${\mathbf H}_D$ is the Dirac Hamiltonian for the unperturbed
Hydrogen atom and ${\mathbf H}^{(n=1)}$ is the perturbation. Also $\boldsymbol\alpha,\boldsymbol\beta$ are
Dirac's matrices. For the solutions to the Dirac equation in a Coulomb potential with energy $E$ and angular
momentum quantum numbers ($\kappa \bar m$) we take the following form \cite{Johnson:2004}
\begin{equation}
\phi_{\bar{n}\kappa \bar m}=\frac{1}{r}\begin{pmatrix}
G_{\bar n\kappa \bar m}(r)\Omega_{\kappa,\bar m}(\vartheta,\varphi)\\
iF_{\bar n\kappa \bar m}(r)\Omega_{\kappa,\bar m}(\vartheta,\varphi)
\end{pmatrix}.
\end{equation}
Here $\bar n=0,1,\dots$ is the principal quantum number. In the Coulomb case the functions $F$ and $G$ are
related to the  confluent hypergeometric functions and $\Omega_{\kappa \bar m}(\vartheta,\varphi)$ are the
spherical spinors. The latter are eigenfunctions of $K=-1-\vec\sigma\cdot \vec L$ with eigenvalues $\kappa=\mp
(j+1/2)$ for $j=l\pm 1/2$. They satisfy the orthogonality relations
\begin{equation}
\int_0^{\pi}\sin\vartheta d\vartheta\int_0^{2\pi}d\varphi\Omega^{\dag}_{\kappa' \bar m'} \Omega_{\kappa \bar
m}=\delta_{\kappa',\kappa}\delta_{\bar m',\bar m}.
\end{equation}
The corresponding energy eigenvalues are
\begin{equation}
E_{\bar n\kappa}=\frac{m}{\sqrt{1+\cfrac{\alpha^2}{(\gamma+\bar n-|\kappa|)^2}}}\;,\;\;
\gamma=\sqrt{\kappa^2-\alpha^2}\,.
\end{equation}
When calculating the energy shifts to lowest order in $\alpha$ the contribution of $F$ can be neglected. The
energy shifts are determined by the matrix elements of the perturbation
\begin{equation}\label{ecu:deltae}
H_{(\bar n\kappa \bar m)(\bar n'\kappa'\bar m')}=\int drr^2d\Omega\phi_{\bar n\kappa \bar m}^{\dag}{\mathbf
H}\phi_{\bar n'\kappa' \bar m'}.
\end{equation}
We want to calculate the energy split between the $2S_{1/2}$ and $2P_{1/2}$ states, $\Delta E^{(n)}:=\delta
E^{(n)}_{2S_{1/2}}-\delta E^{(n)}_{2P_{1/2}}$, due to the presence of the $n$ compact extra dimensions and
compare it with the experimental value for the Lamb shift $\Delta E^{\mathrm{Lamb}}:=\delta
E^{\mathrm{QED}}_{2S_{1/2}}-\delta E^{\mathrm{QED}}_{2P_{1/2}}$ associated to QED.

For the $2S_{1/2}$ state, we have $\bar n=2$ and $j=1/2$, and $G$ is given by
\begin{eqnarray}
G_{2S_{1/2}}(r)&=&-N_{2,-1}\sqrt{1+\sqrt{\frac{\gamma+1}{2}}}\, e^{-m\alpha \sqrt{2(1+\gamma)}\,r} \left(2\alpha
m\sqrt{2(1+\gamma)}\,r\right)^{\gamma}
\nonumber\\
&&\left[(\sqrt{2(1+\gamma)}+1) \left(1-\frac{2m\alpha\sqrt{2(1+\gamma)}\,r}{2\gamma+1}\right)-1\right],
\end{eqnarray}
where
\begin{equation}
N_{2,-1}=\frac{1}{\sqrt{2(1+\gamma)}\Gamma(2\gamma+1)} \sqrt{\frac{\alpha
m\Gamma(2\gamma+2)}{2(\sqrt{2(1+\gamma)}+1)}}.
\end{equation}
The energy shift for $n=1$ determined through  Eq.(\ref{ecu:deltae}) is
\begin{eqnarray} \delta E^{(n=1)}_{2S_{1/2}}&=&\frac{4\alpha\epsilon^3}{3\pi}
\left({2\alpha m}\sqrt{2(\gamma+1)} \right)^{3}N_{2,-1}^2\left[
\left(\frac{\sqrt{2(1+\gamma)}+1}{2\gamma+1}\right)^2\Gamma(2\gamma-1)
\right.\\
&&\left.-\frac{2(\sqrt{2(1+\gamma)}+1)}{2\gamma+1}\sqrt{2(1+\gamma)} \Gamma(2\gamma-2)+
2(1+\gamma)\Gamma(2\gamma-3)\right].\nonumber
\end{eqnarray}
For the $2P_{1/2}$ state $G(r)$ is given by
\begin{eqnarray}
G_{2P_{1/2}}(r)&=&-N_{2,1}\sqrt{1+\sqrt{\frac{\gamma+1}{2}}}\, e^{-m\alpha
\sqrt{2(1+\gamma)}\,r}\,\left(2m\alpha
\sqrt{2(1+\gamma)}\,r\right)^{\gamma}\nonumber\\
&&\left[(\sqrt{2(1+\gamma)}+1)\left(1-\frac{2m\alpha \sqrt{2(1+\gamma)}\,r}{2\gamma+1}\right)+1\right]\,,
\end{eqnarray}
where
\begin{equation}
N_{2,1}=\frac{1}{\sqrt{2(1+\gamma)}\Gamma(2\gamma+1)} \sqrt{\frac{\alpha
m\Gamma(2\gamma+2)}{2(\sqrt{2(1+\gamma)}-1)}}
\end{equation}
and now the energy shift is given by
\begin{eqnarray}
\delta E^{(n=1)}_{2P_{1/2}}&=&\frac{4\alpha\epsilon^3}{3\pi} \left(2\alpha
m\sqrt{2(\gamma+1)}\right)^{3}N_{2,1}^2 \left[ \left(\frac{\sqrt{2(1+\gamma)}+1}{2\gamma+1}
\right)^2\Gamma(2\gamma-1)\right.\\
&&\left.-\frac{2(\sqrt{2(1+\gamma)}+1)}{2\gamma+1} [\sqrt{2(1+\gamma)}+2]\Gamma(2\gamma-2)+
[\sqrt{2(1+\gamma)}+2]^2\Gamma(2\gamma-3)\right].\nonumber
\end{eqnarray}

Next we proceed to consider the case of two compact extra dimensions, $n=2$. Let us take the electrostatic
potential Eq.( \ref{ecu:po2}) and  expand it around Coulomb's, {\em ie} $r>\!\!>\epsilon$, thus producing
\begin{equation}
e^{(n=2)}A_0\approx-\alpha\left\{\frac{1}{r}-\frac{3}{8} \frac{\epsilon^4}{r^5}\right\}\,.\label{eq:eneff}
\end{equation}
Notice that in this case $\alpha:=e^{(n=2)}_{{\mathrm{ eff}}}e^{(n=2)}$. We have now the perturbation
\begin{equation}
{\mathbf H}^{(n=2)}=\frac{3e^2}{8}\frac{\epsilon^4}{r^5}.
\end{equation}
The energy shift for the $2S_{1/2}$ state one arrives at is
\begin{eqnarray}
\delta E^{(n=2)}_{2S_{1/2}}&=&\frac{3e^2\epsilon^4}{8}\left(2\alpha m\sqrt{2(\gamma+1)}
\right)^{4}N_{2,-1}^2\left[ \left(\frac{\sqrt{2(1+\gamma)}+1}{2\gamma+1}\right)^2\Gamma(2\gamma-2)
\right.\\
&&\left.-\frac{2(\sqrt{2(1+\gamma)}+1)}{2\gamma+1}\sqrt{2(1+\gamma)} \Gamma(2\gamma-3)+
2(1+\gamma)\Gamma(2\gamma-4)\right].\nonumber
\end{eqnarray}
For the $2P_{1/2}$ state the corresponding energy shift is
\begin{eqnarray}
\delta E^{(n=2)}_{2P_{1/2}}&=&\frac{3e^2\epsilon^4}{8}\left(2\alpha m\sqrt{2(\gamma+1)}
\right)^{4}N_{2,1}^2\left[ \left(\frac{\sqrt{2(1+\gamma)}+1}{2\gamma+1}\right)^2\Gamma(2\gamma-2)
\right.\\
&&\left.-\frac{2(\sqrt{2(1+\gamma)}+1)}{2\gamma+1}[\sqrt{2(1+\gamma)}+2]
\Gamma(2\gamma-3)+[\sqrt{2(1+\gamma)}+2]^2\Gamma(2\gamma-4)\right] .\nonumber
\end{eqnarray}
Now we are ready to compare the $2S_{1/2}$-$2P_{1/2}$ energy shift for Hydrogen in the presence of $n$ compact
and one warped extra dimensions with the QED Lamb shift. Namely, by assuming a null result from the experiment
that could be directly associated to the extra dimensions we have \cite{Johnson:2004,Itzykson:1980rh}
\begin{equation}
\Delta E^{(n)} < \Delta E^{\mathrm{Lamb}}\,,
\end{equation}
and with the use of the explicit formulae obtained above we can estimate bounds for $\epsilon$. Notice this has
the advantage with respect the use for comparison of the uncertainty in the measurement that one does not need
to consider other important effects as the charge radius of the proton, recoil effect, among others, which are
indeed less than $\Delta E^{\mathrm{Lamb}}$ \cite{MohrTaylor:2005}. Using the value of $\Delta
E^{\mathrm{Lamb}}/h=1057845.0(9.0)kHz$ \cite{MohrTaylor:2005} the following results follow: $\epsilon <
10^{-14}m,~ 10^{-13}m$ for $n=1,2$, respectively.

\section{Discussion}

In this work we have investigated low-energy effects associated to the presence of extra dimensions. The brane
world considered here is a Randall-Sundrum like containing a $(3+n)$-brane embedded in a $(3+n+1)$ space with
$n$ compact  and one warped infinite extra dimensions or RSIIn. Apart from its hybrid character having both
compact and non-compact extra dimensions this model is interesting because it localizes the zero mode gauge
field due to the presence of the compact dimensions \cite{Dubovsky:2000av}. The effects we have considered are
i) the effective electromagnetic potentials of a point charge moving uniformly  along the non compact directions
of the $(3+n)$-brane and ii) the Lamb shift for a Hydrogen atom localized in the same subspace.

We took a test point charge as seen from the 4D spacetime perspective, moving uniformly along a timelike
straight geodesic spatially contained along the non compact directions of the $(3+n)$-brane. Some comments
regarding the stability of such geodesic are in order. Although unstable such geodesics are interesting whenever
the tunneling time into the extra dimensions is large enough. As it is clear from our discussion above for the
spinor localization based on \cite{Dubovsky:2000am} for RSIIn this can be the case quantum mechanically. It has
also been proved for other similar scenarios in \cite{Dubovskytunnel}. Thus it is reasonable to think in terms
of a semiclassical approximation leading to such a tunneling effect with long life-time for our uniformly moving
particle. We assume then that our description of the classical test charge is the leading order of the
corresponding semiclassical approximation for the quantum situation. Moreover, at high enough energies, a
particle gets excited along the compact extra dimensions and in this case the instability of the geodesic is
removed.

Our results summarize as follows. At low energies our point charge in uniform motion produces electromagnetic
potentials given by the zero mode regarding the compact extra dimensions but include light massive modes
associated to the non-compact dimension. They reduce to the standard Liennard-Wiechert potentials in 4D
Minkowski spacetime at long distances, as determined by the characteristic scale of the scenario $\epsilon$.
Remarkably below this scale they differ from the standard ones and become finite at the position of the point
charge. They can easily be seen to produce finite self energies for the point particles as calculated in 4D
namely $E_{\mathrm{self}}^{(n=1)}=\frac{5}{32\pi^3}\frac{(e^{(n=1)}_{\mathrm{eff}})^2}{\epsilon}$ and
$E_{\mathrm{self}}^{(n=2)}=\frac{315}{16384\pi}\frac{(e^{(n=2)}_{\mathrm{eff}})^2}{\epsilon}$. The details of
the potentials for short distances depend on the number of extra dimensions $n$. It is tempting to explain the
finiteness of the electromagnetic potentials and self energy due to the fact that our charge looks point-like in
3D but it is actually effectively smeared along the compact extra dimensions at low energies due to the zero
mode approximation in the compact dimensions for the electromagnetic fields. Indeed as $R\rightarrow 0$ in
(\ref{ecu:po1}) and (\ref{ecu:po2}) such finiteness is lost. However it must be observed that for  finite $R$
actually the standard divergences for the Coulomb potential and self-energy are regained in the limit $\epsilon
\rightarrow 0$. So the finiteness is better described as a combined effect of the presence of the compact and
warped extra dimensions. Indeed not having compact dimensions amounts to an impossibility of localizing the
particle's electromagnetic field to the brane.

As for the Hydrogen atom model we take a  Dirac equation in RSIIn for which we have adopted the localization
mechanism of the spinor as given in \cite{RD-Sh,Dubovsky:2000am,Neronov:2001qv}. Here we pick the zero mode in
both compact and non-compact extra dimensions for the localized spinor and couple it with the static form of the
time component of the effective Liennard-Wiechert potential we have determined in the present work, Eqs.
(\ref{ecu:po1}) and (\ref{ecu:po2}). For the static case the effective potential can be seen as differing from
the standard 4D Coulomb potential by an additive term containing a different power of the spatial separation
between the point like charges. The power depends upon the number $n$ of compact extra dimensions. By making use
of a perturbation analysis in terms of this deviation from the Coulomb potential it was possible to calculate a
$2S_{1/2}-2P_{1/2}$ shift which we compared with experimental Lamb shift data $\Delta
E^{\mathrm{Lamb}}/h=1057845.0(9.0)kHz$ \cite{MohrTaylor:2005}. Assuming the effect of the extra dimensions
$\Delta E^{(n)}$ is smaller than the Lambs's $\Delta E^{\mathrm{Lamb}}$ yields the bounds to the length scale of
the scenario: $\epsilon < 10^{-14}m,~ 10^{-13}m$ for $n=1,2$, respectively. Notice we estimated such bounds  by
comparing it with the Lamb shift itself and not with the experimental uncertainty in its measurement. So no
other effects like the charge radius of the proton or the recoil effect, among others, are relevant since they
are indeed less than $\Delta E^{\mathrm{Lamb}}$. Of course one could insist on using such uncertainty but no
significant improvement is obtained because although the bounds improve one has then to incorporate the afore
mentioned extra effects that essentially cancel the improvement. Furthermore our upper bounds for $\epsilon$ lie
in the fermi range so no conflict with macroscopic electromagnetism tests is expected. To go beyond such scale
say scattering experiments one would actually require a quantum field approach or QED on RSIIn which remains an
interesting open problem.

Overall both of our results add evidence on how known low-energy phenomena can be used to probe models with
extra dimensions that were originally thought to be testable only at high energies while being automatically
compatible with standard low-energy physics.

Remarkably, finiteness of the electromagnetic potentials and self-energy has led in the past to new ideas as for
instance the proposal that gravity regulates the self-energy of a point particle \cite{ADM} or a non linear
electrodynamics of Born and Infeld \cite{BI} implementing the finiteness of the electric field \cite{BI}. In
light of our results it would be of interest to study other singular problems in standard electrodynamics as the
radiation reaction problem. In this regard higher dimensional flat spacetimes have been considered
\cite{Galtsov1, Galtsov2} but curved hybrid scenarios like RSIIn have not to the best of our knowledge. As for
the quantum extensions it would be interesting to study further the finite character we have found here for the
potentials and self-energy of a point charge quantum mechanically and its possible bearing on QED in particular
high precision tests \cite{precision-QED} or possibly other gauge fields of the standard model \cite{QED-SM,
colliderEXP} to see whether they may add something to realistic brane world models \cite{kokorelis, Dbrane}.

This work was partially supported by Mexico's National Council of Science and Technology (CONACyT), under grants
CONACyT 51132-F and CONACyT-NSF E-120-0837. O.P. acknowledges the support of the CONACyT fellowship 162767. He
would also like to thank the Young collaborator Programme of Abdus Salam ICTP Trieste, Italy, for supporting a
visit where part of this work was done.


\end{document}